\documentclass[%
 reprint,
 superscriptaddress,
 amsmath,amssymb,
 aps,
 pra,
]{revtex4-1}

\usepackage{graphicx}
\usepackage{dcolumn}
\usepackage{bm}
\usepackage[hyperfootnotes=true,
            colorlinks=true,
            linkcolor=blue,
            anchorcolor=blue,
            citecolor=blue,
            urlcolor=blue
            ]{hyperref}

\usepackage[utf8]{inputenc}
\usepackage[T1]{fontenc}
\usepackage{mathptmx}

\usepackage[english]{babel}
\usepackage{indentfirst}
\usepackage{graphicx}
\usepackage{subfigure}
\usepackage{algorithm}
\usepackage{algorithmicx}
\usepackage{algpseudocode}                        
\usepackage{amsmath}
\usepackage{amssymb}
\usepackage{caption}
\usepackage{enumerate}
\usepackage{appendix}
\usepackage{ragged2e}
\justifying\let\raggedright\justifying            

\begin{document}

\preprint{APS/123-QED}
\captionsetup[figure]{labelformat={default},labelsep=period,name={FIG.},singlelinecheck=off,justification=raggedright}

\title{Multi-matrix rate-compatible reconciliation for quantum key distribution}

\author{Chao-hui Gao}
 \thanks{These authors contributed equally to this work.}
\author{Yu Guo}
 \thanks{These authors contributed equally to this work.}
\author{Dong Jiang}%
 \thanks{jiangd@nju.edu.cn}
\author{Li-jun Chen}
 \thanks{chenlj@nju.edu.cn}
 \address{State Key Laboratory for Novel Software Technology, Nanjing University, Nanjing, 210046, P.R.China}

\date{\today}

\begin{abstract}
Key reconciliation of quantum key distribution (QKD) is the process of correcting errors caused by channel noise and eavesdropper to identify the keys of two legitimate users.
Reconciliation efficiency is the most important figure for judging the quality of a reconciliation scheme.
To improve reconciliation efficiency,
rate-compatible technologies was proposed for key reconciliation,
which is denoted as the single-matrix rate-compatible reconciliation (SRCR).
In this paper,
a recently suggested technique called multi-matrix reconciliation is introduced into SRCR,
which is referred to as the multi-matrix rate-compatible reconciliation (MRCR),
to further improve reconciliation efficiency and promote the throughput of SRCR.
Simulation results show that MRCR we proposed outperforms SRCR in reconciliation efficiency and throughput.
\end{abstract}

\maketitle


\section{\label{sec:level1}Introduction}

Quantum computing possesses a threat on conventional cryptographic tools \cite{RivestA} based on computational complexity \cite{shor1999polynomial}.
In these circumstances,
quantum key distribution (QKD) promises unconditional security guaranteed by laws of quantum mechanics \cite{scarani2009security}.
Therefore, it has attracted widespread attention \cite{lo2005decoy,Wang2005Beating,lo2012measurement,liao2017satellite} last decades, and is currently being deployed in commercial applications \cite{sasaki2011field,courtland2016china}.

QKD can realize secure key distribution between two legitimate users, i.e. Alice and Bob, even when eavesdropper Eve is present.
However, because of the noise in quantum channel and the existence of Eve, there are some errors in Alice's and Bob's keys.
To cope with this problem,
key reconciliation is introduced into QKD as the process of correcting errors to identify Alice's and Bob's keys, and is performed via some algorithms such as Belief Propagation (BP) \cite{mceliece1998turbo,kschischang2001factor} and Log Likelihood Ratio BP (LLR-BP) \cite{chung2001analysis,liu2014variable}, which is the log version of BP and  can reduce plenty of computation, as such is widely used.
For convenience, we refer BP and LLR-BP as the single-matrix reconciliation (SR).

SR corrects errors in keys by limited iterations with the help of syndrome \cite{richardson2001capacity} sent by Alice and low-density parity-check (LDPC) code
which was proposed by Gallager \cite{Gallager1962Low} in 1962.
LDPC code is a linear block error correction code,
and can be represented by a $m*n$ binary spare matrix,
a row and a column of which correspond to a check node $c_j\ (j\in \{1,\ 2,\ \cdots,\ m\})$ and a variable node $v_i\ (i\in \{1,\ 2,\ \cdots ,\ n\})$, respectively.
In a matrix $H_{m*n}$, we denote the value in Row $j$ and Column $i$ by $H_{ji}$.
And the set of the neighboring check nodes of $v_i$ is defined by $\{c_j|H_{ji}=1\}$,
while the set of the neighboring variable nodes of $v_i$ is $\{v_k|H_{jk}=1,\ H_{ji}=1\ and\ v_k\neq v_i\}$
which has $N(v_i)$ elements.
LDPC code plays an important role in SR.
When the reconciliation begins,
Alice makes use of her key and the matrix shared with Bob to calculate the syndrome \cite{richardson2001capacity} and sends the syndrome to Bob.
Then Bob implements SR to correct errors from his key in conjunction with LDPC code and syndrome.
The detail can be found in \cite{chung2001analysis,liu2014variable}.

However, SR gradually cannot satisfy the requirements of reconciliation efficiency with the development of QKD systems.
Thus, rate-compatible technologies, i.e. shortening and puncturing, were introduced into SR
(hereinafter referred to as the single-matrix rate-compatible reconciliation, SRCR).
Unfortunately, SRCR shows a poor performance in throughput, though the improvement in reconciliation efficiency.

In \cite{gao2019multi}, a LLR-BP based reconciliation scheme using multiple matrices (or multi-matrix reconciliation, MR) to correct errors was proposed, which greatly improves the throughput by increasing the convergence speed.
In this paper, we introduce MR technology into SRCR to optimize reconciliation efficiency further
and promote the throughput compared with SRCR.
To verify our views, we perform several numerical simulations and demonstrate the huge advantages in reconciliation efficiency and throughput of our scheme, which provides a promising way to improve the secure key generation rate in QKD systems.

The rest of the paper is organized as follows: In section II, we review the LDPC code and the process of key reconciliation first. Then rate-compatible technologies and the concepts of SRCR are introduced. In section III, we depict our scheme, i.e. multi-matrix rate-compatible reconciliation (MRCR) in detail. Section IV gives the performance evaluations of the proposed scheme and SRCR.
Finally, the conclusions are presented in Section V.

\section{\label{sec:level2}Single-matrix Rate-compatible Reconciliation}

There is an important figure called the reconciliation efficiency $f$ \cite{kiktenko2018error,kiktenko2017symmetric,elkouss2009efficient}, which shows the ratio of the amount of information published during reconciliation to the theoretical minimum amount of information necessary for successful reconciliation.
Thus, to correct errors successfully, $f>1$ must be hold in an effective reconciliation scheme.
For single-matrix reconciliation (SR) and MR \cite{gao2019multi}, $f$ can be represented as
\begin{equation}
f=\frac{m}{nh(e)}=\frac{1-R_0}{h(e)}>1,
\label{equ:efficiency_f}
\end{equation}
where $m$ and $n$ are numbers of rows and columns of the LDPC codes,
$R_0=1-m/n$ is the initial code rate,
$e$ is the result of error estimation,
and $h(e)$ is the Shannon binary entropy of $e$:
\begin{equation}
h(e)=-elog_2e-(1-e)log_2(1-e).
\end{equation}
Reconciliation efficiency $f$ is used to characterize security of a reconciliation scheme,
and to remove information leakage as the key figure during privacy amplification \cite{bennett1988privacy,bennett1995generalized}.
And less information needs to be removed if $f$ is closer to $1$.

For finely tuning $f$ to approach $1$, two rate-compatible techniques known as shortening and puncturing \cite{kiktenko2017symmetric,elkouss2010secure} can be employed to modify $R_0$ in Eq. (\ref{equ:efficiency_f}) by changing Alice's and Bob's sifted keys ($\bm{X}$ and $\bm{Y}$) rather than displacing the LDPC matrix.
When shortening, $s$ shortened bits are published, which is equivalent to converting $R_0$ from $1-m/n$ to $1-m/(n-s)$.
Whereas, when puncturing, Alice and Bob use two independent true random number generators (TRNGs) to produce the values of $p$ punctured bits. The process is equivalent to converting $R_0$ from $1-m/n$ to $1-(m-p)/(n-p)$.
As specified above, the shortening (puncturing) serves for lowering (raising) $R_0$.
If Alice and Bob perform key reconciliation successfully, they remove $s$ shortened bits and $p$ punctured bits to obtain corrected sifted key,
and $f$ is finely tuned to the following form:
\begin{equation}
f=\frac{m-p}{(n-p-s)h(e)}=\frac{1-R_1}{h(e)}>1,
\label{equ:new_f}
\end{equation}
where $R_1=1-(m-p)/(n-p-s)$ is the adjusted code rate.

The positions of shortened bits could be chosen from punctured bits or via a pseudo random number generator (PRNG) without compromising the performance.
However, for puncturing, there are theoretical and experimental studies \cite{ha2004rate,hsu2008capacity} show that the positions chosen intentionally outperform the ones from PRNG.
In the intentionally puncturing algorithms \cite{elkouss2012untainted,ha2006rate,vellambi2009finite}, the untainted-puncturing algorithm (UPA) \cite{elkouss2012untainted} is simpler, more efficient and thus more popular than others.
UPA chooses punctured bits that avoid the generation of dead check nodes, each of which is connected with at least two punctured bits.
Dead check nodes erase reconciliation information and significantly degrade the performance of reconciliation,
and that is why UPA outperforms others.
The process of UPA is as follows:

\begin{algorithm}[H]
\caption{Untainted-puncturing Algorithm (UPA)}
\begin{algorithmic}[1]                            
    \State $\Omega \longleftarrow \{v_1, v_2, \cdots, v_n\}$
    \State Initialize $N(v_i)\ (i\in \{1, 2, \cdots, n\})$
    \While{$\Omega \ne \emptyset$}
        \State Find $\bar{v}\in\Omega\ \big(N(\bar{v})\leq\ N(\tilde{v}), \forall \tilde{v}\in\Omega\big)$
        \State $N_{min} \longleftarrow N(\bar{v})$
        \State $\bar{\Omega} \longleftarrow \emptyset$
        \For{every $\tilde{v}\in\Omega$}
            \If{$N(\tilde{v})=N_{min}$}
                \State Add $\tilde{v}$ to $\bar{\Omega}$
            \EndIf
        \EndFor
        \State Choose $v_p$ randomly from $\bar{\Omega}$ as a punctured bit
        \State Delete $v_p$ and its neighboring variable nodes from $\Omega$
    \EndWhile
\end{algorithmic}
\end{algorithm}


After the proposal for shortening, puncturing and UPA,
they are applied to SRCR \cite{martinez2013key}.
In SRCR, the range of $e$ that is suitable for reconciliation is divided into several intervals,
and each interval corresponds to a initial code rate.
In other words, the initial code rate, $R_0=1-m/n$, is determined by which interval $e$ in.
Then Alice and Bob derive the numbers of initial punctured and shortened bits needed to achieve the desired reconciliation efficiency $f_d$ as follows \cite{kiktenko2017symmetric}:
\begin{equation}
\label{equ:ps_initial}
\begin{aligned}
p_0&=\lfloor \frac{m-nh(e)f_d}{1-h(e)f_d} \rfloor , \\
s_0&=0.
\end{aligned}
\end{equation}
The first $p_0$ results of UPA are chosen as punctured bits, and respectively assigned true random values by the parties.
UPA usually produces far more than $p_0$ bits, otherwise the parties jointly decide the rest of punctured bits via PRNG.
Next, they exchange syndromes \cite{richardson2001capacity} based on their punctured sifted keys ($\bm{X_p}$ and $\bm{Y_p}$).
Bob decodes $\bm{Y_p}$ via LLR-BP \cite{gao2019multi,kiktenko2017symmetric,chung2001analysis}.
The steps up to the present are called a communication round.
If Bob fails to decode $\bm{Y_p}$, the parties enter the next communication round and
Alice reduces current code rate by shortening, i.e., publishing values of some punctured bits,
to increase the probability of successful reconciliation.
Bob corrects the values of these shortened bits in $\bm{Y_p}$, and then decodes it again.
The process will come to an end when Bob finds the correct sifted key or all of $p_0$ punctured bits have been revealed as shortened bits.

\section{\label{sec:level3}Multi-matrix Rate-compatible Reconciliation}

Besides reconciliation efficiency, throughput is another important figure,
which tells the number of key bits processed per unit of time.
Throughput can be noticeably improved in a reconciliation algorithm with faster convergence speed.
With this motivation, a multi-matrix reconciliation technique, i.e. MR, has been proposed in \cite{gao2019multi},
where in each iteration multiple matrices produce more useful information to correct errors
such that the iteration number falls and the convergence speed increases.
Further experiments reveal that the technique can achieve higher success rate,
since cycles \cite{yazdani2004improving}, which appear in one matrix and can degrade the performance of the matrix,
can be weakened by other matrices to avoid reconciliation failures.

In this letter, we introduce MR technique into SRCR,
and refer to this combination as the multi-matrix rate-compatible reconciliation (MRCR).
In a communication round, MRCR can provide more timely information to assist Bob in decoding $Y_p$
and reduce interference from cycles,
thereby achieving higher success rate compared with SRCR.
In other words, it takes fewer communication rounds to complete reconciliation in MRCR.
And this means not only the increase in convergence speed,
but also the improvement in reconciliation efficiency $f$ because of the decrease in the number of shortened bits.

In MRCR, we first consider choosing punctured bits via UPA.
However, the chosen punctured bits in one matrix are very likely to generate dead check nodes in other matrices.
To solve the problem, only the key bits which appear in all the results can be chosen for puncturing
after we run UPA for each matrix.
Investigations further show that the number of punctured bits obtained by this solution is far from
the number required to achieve the desired reconciliation efficiency $f_d$.
Thus, we allow dead check nodes to exist,
but their number should be kept as small as possible.
In this way, we have improved UPA and refer to this UPA-based algorithm for MRCR
as the multi-matrix untainted-puncturing algorithm (MUPA), which is presented below.

\begin{algorithm}[H]
\caption{Multi-matrix Untainted-puncturing Algorithm (MUPA)}
\begin{algorithmic}[1]                            
    \State $\bar{p} \longleftarrow\ 0$
    \State $k \longleftarrow\ 0$
    \State Initialize $N(v_i)\ (i\in \{1, 2, \cdots, n\})$
    \While{$\bar{p}<p_0$}
        \State Initialize $\Omega_k$, which is the set of variable nodes have $k$ neighboring punctured bits in all of the $N$ matrices
        \While{$\Omega_k \ne \emptyset$ and $\bar{p}<p_0$}
            \State Find $\bar{v}\in\Omega_k\ \big(N(\bar{v})\leq\ N(\tilde{v}), \forall \tilde{v}\in\Omega_k\big)$
            \State $N_{min} \longleftarrow N(\bar{v})$
            \State $\bar{\Omega} \longleftarrow \emptyset$
            \For{every $\tilde{v}\in\Omega_k$}
                \If{$N(\tilde{v})=N_{min}$}
                    \State Add $\tilde{v}$ to $\bar{\Omega}$
                \EndIf
            \EndFor
            \State Choose $v_p$ randomly from $\bar{\Omega}$ as a punctured bit
            \State Delete $v_p$ and its neighboring variable nodes from $\Omega_k$
            \State $\bar{p} \longleftarrow\ \bar{p}+1$
        \EndWhile
        \State $k \longleftarrow\ k+1$
    \EndWhile
\end{algorithmic}
\end{algorithm}

Alice and Bob take advantage of MUPA to select the initial $p_0$ punctured bits during MRCR,
before which the parties should decide $N$, the number of matrices participate in reconciliation,
$U_L$, a positive integer which can prevent MRCR from falling into an infinite loop,
and $\delta<1$, the ratio of newly generated shortened bits to $p_0$ after a communication round.
Then, MRCR is performed as follows:

\begin{enumerate}[1.\itemindent=-\itemindent]
\item We assume that the estimated error rate $e$ has been derived from error estimation \cite{Wang2005Beating,kiktenko2018error}.
Similarly to SRCR, the range of bit error rate (BER) suitable for reconciliation is split into several intervals characterized by initial code rates.
So the initial code rate $R_0$ is determined by which interval $e$ in.
After that, $N$ matrices, $H_1,\ H_2,\ \cdots,\ H_N$, with code rate $R_0$ are decided by the parties.
Then, the numbers $p_0$ and set $\mathbb{P}$ of initial punctured bits are given by Eq. (\ref{equ:ps_initial}) and MUPA respectively,
whereas the set $\mathbb{S}$ of initial shortened bits is empty.

\item According to the positions of initial punctured bits,
The parties make use of TRNGs to modify their sifted keys ($\bm{X}$ and $\bm{Y}$) to obtain the punctured sifted keys ($\bm{X_p}$ and $\bm{Y_p}$).
Afterwards Alice calculates $N$ syndromes, $\bm{Z_1},\ \bm{Z_2},\ \cdots,\ \bm{Z_N}$, via Eq. (\ref{equ:Alice_syndrome})
and transmits them to Bob through the classical authenticated channel.
\begin{equation}
\label{equ:Alice_syndrome}
\begin{aligned}
(\bm{Z_1})^{T}&=(H_1\cdot \bm{X_p}^{T})\pmod{2}, \\
(\bm{Z_2})^{T}&=(H_2\cdot \bm{X_p}^{T})\pmod{2}, \\
&\cdots \\
(\bm{Z_N})^{T}&=(H_N\cdot \bm{X_p}^{T})\pmod{2}.
\end{aligned}
\end{equation}

\item Bob decodes $\bm{Y_p}=\left[Y_p^1,\ Y_p^2,\ \cdots,\ Y_p^n \right]$ mainly based on MR \cite{gao2019multi},
    where step $3.2\sim3.5$ is called an iteration.
    \begin{enumerate}[3.1.\itemindent=1em]
    \item Initialize the prior probabilities ${P_r}^{0}_i$, ${P_r}^{1}_i$, log likelihood ratios ${Log}_{i}$  and variable-to-check (V2C) information $I_{v_i\to c_j}^k$ as below:
        \begin{equation}
        \begin{cases}
        {P_r}^{0}_i=1-e, {P_r}^{1}_i=e & (Y_p^i=0)\\
        {P_r}^{1}_i=1-e, {P_r}^{0}_i=e & (Y_p^i=1)
        \end{cases},
        \label{equ:bp_initial}
        \end{equation}

        \begin{equation}
        {Log}_{i}=
        \begin{cases}
        \log \frac{{P_r}^{0}_i}{{P_r}^{1}_i} & (v_i\notin \mathbb{P}\ and\ v_i\notin \mathbb{S})\\
        0 & (v_i\in \mathbb{P})\\
        +\infty & (v_i\in \mathbb{S}\ and\ Y_p^i=0)\\
        -\infty & (v_i\in \mathbb{S}\ and\ Y_p^i=1)
        \end{cases},
        \label{equ:L-Pi}
        \end{equation}

        \begin{equation}
        I_{v_i\to c_j}^k={Log}_{i},
        \label{equ:init-Lq}
        \end{equation}
        where $i\in \{1, 2, \cdots, n\}$, $j\in \{1, 2, \cdots, m\}$, $k\in \{1, 2, \cdots, N\}$.

    \item Generate check-to-variable (C2V) information $I_{c_j\to v_i}^k$ as follows,
        \begin{equation}
        \begin{aligned}
        I_{c_j\to v_i}^k=&2\cdot \mathrm{Sign}\left(Z_j^k\right)\cdot \mathrm{tanh}^{-1}\cdot \\
        &\left(\prod_{v_{i'}\in{{Ne}^k\left(c_j\right)\backslash i}}\mathrm{tanh}\left(\frac{1}{2}I_{v_{i'}\to c_j}^k\right)\right),
        \label{equ:L-rji-old}
        \end{aligned}
        \end{equation}
        where $Z_j^k$ is the $j^{th}$ bit of Alice's syndrome $\bm{Z_k}$,
        $Sign$ is a signum function which outputs $+1$ (or $-1$) when $Z_j^k=0$ (or $1$),
        $v_{i'}\in{{Ne}^k\left(c_j\right)\backslash i}$ represents any neighboring variable node of $c_j$ except $v_i$ in $H_k$,
        $tanh$ is the hyperbolic tangent function with its inverse function $tanh^{-1}$.
        If $c_j$ is a dead check node in $H_k$,
        then in the first iteration, for $\forall v_i\in {Ne}^k(c_j)$,
        a punctured bit $v_{i'}\in {Ne}^k(c_j)\backslash i$ with $I_{v_{i'}\to c_j}^k={Log}_{i}=0$ always exists
        and leads to $I_{c_j\to v_i}^k=0$,
        which means all of this C2V information is wiped away.
        Fortunately, the number of dead check nodes has been minimized by MUPA and will decrease when punctured bits turn into shortened bits.
        Additionally dead check nodes can erase C2V information only in the first iteration,
        so in MRCR the impact of dead check nodes is very limited,
        which is confirmed in the subsequent experiments.

    \item Generate V2C information $I_{v_i\to c_j}^k$ by
        \begin{equation}
        I_{v_i\to c_j}^k={Log}_i+\sum_{c_{j^{'}}\in{{Ne}^k\left(v_i\right)\backslash j}}I_{c_{j^{'}}\to v_i}^k,
        \label{equ:L-qij}
        \end{equation}
        where $c_{j^{'}}\in{{Ne}^k\left(v_i\right)\backslash j}$ represents any neighboring check node of $v_i$ except $c_j$ in $H_k$.

    \item For all the variable nodes except shortened bits, obtain their soft-decision values by
        \begin{equation}
        I_{v_i}={Log}_i+\sum^{N}_{k=1}\sum_{c_j\in{{Ne}^k\left(v_i\right)}}{I^k_{c_j\to v_i}}\ \ (v_i\notin \mathbb{S}),
        \label{equ:M-decision}
        \end{equation}
        and make decoding decisions via
        \begin{equation}
        Y_p^i=
        \begin{cases}
        0& I_{v_i}>0 \\
        1& I_{v_i}<0
        \end{cases}
        \ \ (v_i\notin \mathbb{S}).
        \label{equ:decode-decide}
        \end{equation}

    \item An operation will be used at this stage:
        current error rate $\bar{e}$ is estimated by the method called multi-syndrome error estimation \cite{gao2019multi},
        and compared with error rate $\tilde{e}$ of last iteration.
        If $\bar{e}>\tilde{e}$, restore $\bm{Y_p}$ to the state of last iteration and go to step $4$;
        otherwise, return to step $3.2$ and begin a new iteration.
        We refer to this operation as $comparison\ of\ error\ rates$.

        At this stage, if the following equations are all satisfied, it means $\bm{Y_p}$ has been decoded successfully,
        then the parties exit MRCR;
        \begin{equation}
        \label{equ:Bob_syndrome}
        \begin{aligned}
        (\bm{Z_1})^{T}&=(H_1\cdot \bm{Y_p}^{T})\pmod{2}, \\
        (\bm{Z_2})^{T}&=(H_2\cdot \bm{Y_p}^{T})\pmod{2}, \\
        &\cdots \\
        (\bm{Z_N})^{T}&=(H_N\cdot \bm{Y_p}^{T})\pmod{2}.
        \end{aligned}
        \end{equation}
        if Eqs. (\ref{equ:Bob_syndrome}) are not satisfied and the number of iterations in this communication round is less than upper limit $U_L$,
        then carry out $comparison\ of\ error\ rates$;
        otherwise go to step $4$.
    \end{enumerate}

\item If $|\mathbb{P}|>0$,
    Alice randomly selects $P2S$ punctured bits from $\mathbb{P}$ as below,

    \begin{equation}
    P2S=
    \begin{cases}
    \lfloor p_0\cdot\delta\rfloor & (1\leq p_0\cdot\delta \leq |\mathbb{P}|)\\
    |\mathbb{P}| & (1\leq |\mathbb{P}| < p_0\cdot\delta)\\
    1 & (p_0\cdot\delta <1)
    \end{cases},
    \label{equ:P2S}
    \end{equation}

    change them into shortened bits by publishing their positions and values, put them into $\mathbb{S}$.
    Bob corrects $\bm{Y_p}$ according to the positions and values published,
    then goes back to step $3$ and starts a new communication round.
    Otherwise, the parties exit MRCR and fail to decode $\bm{Y_p}$.
    Throughout the process, $|\mathbb{P}|+|\mathbb{S}|\equiv p_0$.
\end{enumerate}

\section{\label{sec:level4}Experimental Evaluation}

As described previously,
rate-compatible technologies and MR technology are used respectively to enhance the performance of SR in reconciliation efficiency and throughput, which are both significantly improved in the scheme we proposed, i.e. MRCR.
In order to verify it, at first we provide detailed comparisons of reconciliation efficiency in different reconciliation algorithms.
Then the potential of MRCR for achieving superior reconciliation efficiency is evaluated further.
After that, a numerical experiment is carried out to fully exhibit the advantage of MRCR in throughput.

In simulation setups,
we first fix $N$ to $3$ and construct LDPC codes with four widely used code lengths $n=5000$, $10000$, $15000$, and $25000$ \cite{miladinovic2004systematic,jang2012design,cushon2015energy,cushon2016low},
each of which includes three code rates $R_0=0.6,\ 0.7$ and $0.8$
for MRCR by the multi-matrix construction method \cite{gao2019multi} (see Appendix B for details).
And for each $n$ and each $R_0$, a matrix applied to SRCR is randomly selected from $3$ matrices in MRCR.
Next, for each $n$ we generate $500$ sets of keys at each signal-noise ratio (SNR) value,
which is within nineteen SNR values ranging from $3.51dB$ to $7.48dB$.
Additionally, $p_0$ obtained from Eq. (\ref{equ:ps_initial}) is required by UPA and MUPA to decide the positions of punctured bits at any $n$, $R_0$, SNR
mentioned above with $f_d=1.1$, $1.08$, $1.06$ and $1.04$.
And we set $U_L$ and $\delta$ to $100$ and $0.02$ respectively throughout the simulations which are carried out under an additive white Gaussian noise (AWGN) channel.

\subsection{Reconciliation Efficiency}
\begin{figure*}[htbp]
    \centering
    \includegraphics[width=15cm]{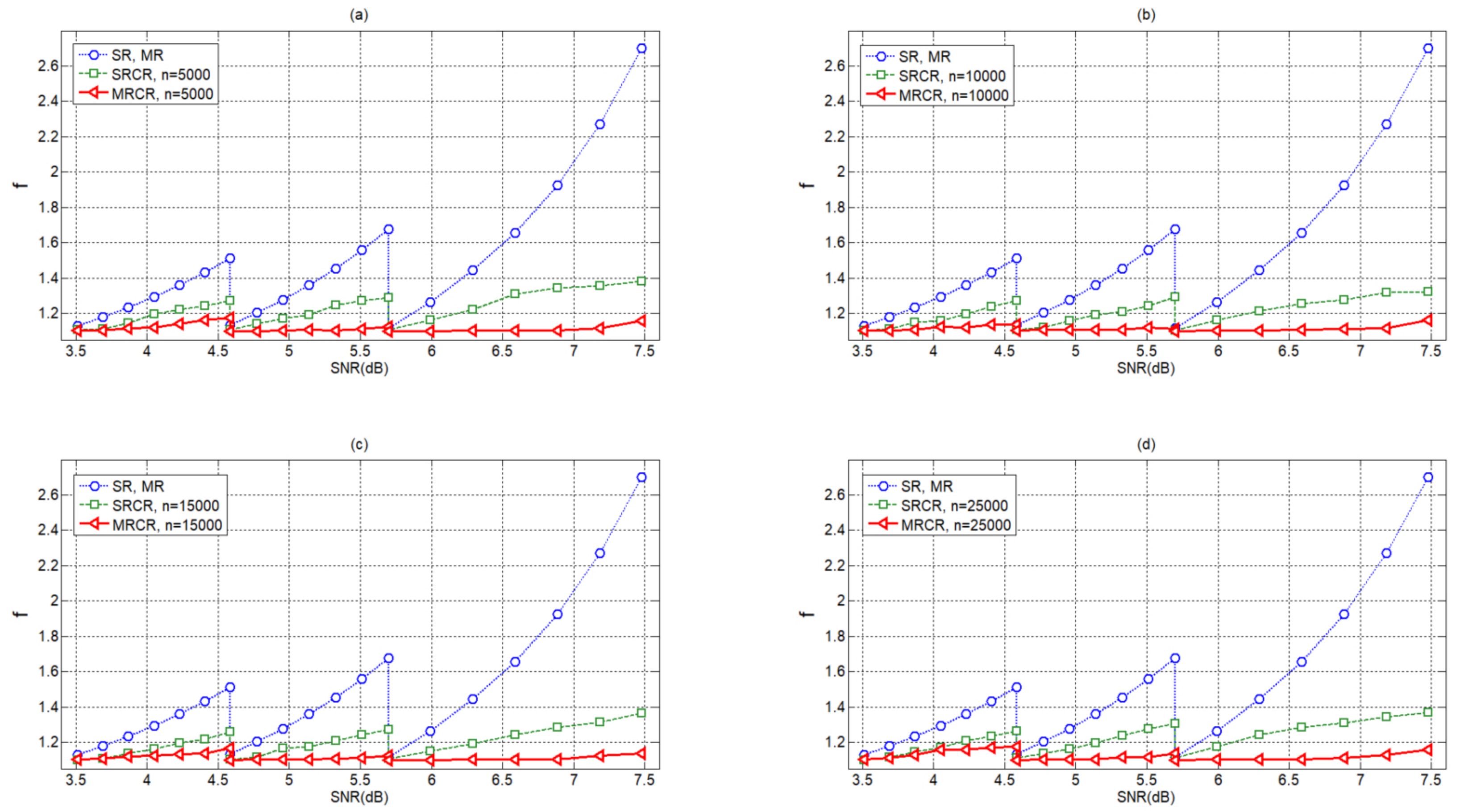}
    \caption{\label{fig:codeLength} Reconciliation efficiency $f$ of SR, MR, SRCR and MRCR are compared as a function of SNR.
    Four different code lengths are considered: (a) $5000$, (b) $10000$, (c) $15000$ and (d) $25000$.}
\end{figure*}

Compared to SRCR,
MRCR can bring faster convergence speed and fewer communication rounds,
leading to the improvement in reconciliation efficiency $f$.
To confirm that, we perform a numerical simulation with $n=5000$, $10000$, $15000$ and $25000$ respectively.
For every SNR in each $n$, $500$ sets of keys were decoded in MRCR,
and the mean of $f$ obtained by Eq. (\ref{equ:new_f}) is calculated for successful reconciliation,
so does SRCR.

As illustrated in Fig. \ref{fig:codeLength},
the red lines are underneath the green lines and the blue lines.
This observation suggests that MRCR can achieve better $f$ than SRCR, SR and MR in any $n$ and any SNR.

In addition, it is clearly seen that the blue lines standing for SR and MR display a saw pattern \cite{elkouss2009efficient},
which comes from the fact that we use the discrete code rates $R_0=0.6$, $0.7$ and $0.8$.
At each code rate, there is a SNR threshold where $f$ performs best.
And according to Eq. (\ref{equ:efficiency_f}), $f$ trends to be worse along with the increase of SNR.
The saw pattern has a negative influence on secure key generation rate because of frequent changes of parameters during privacy amplification.
Obviously, it's impractical to eliminate the saw pattern by implementing continuous code rates.
Fortunately rate-compatible techniques can mitigate the saw pattern.
Referring to Fig. \ref{fig:codeLength},
the saw behavior of green lines representing SRCR is much gentler,
whereas the saws are nearly eliminated on the red lines standing for MRCR.

On balance, MRCR is able to keep $f$ optimal and stable at any $n$, $R_0$ and SNR,
therefore it has positive impact on secure key generation rate.

\subsection{Potential for Better Reconciliation Efficiency}
\begin{figure*}[htbp]
    \centering
    \includegraphics[width=10cm]{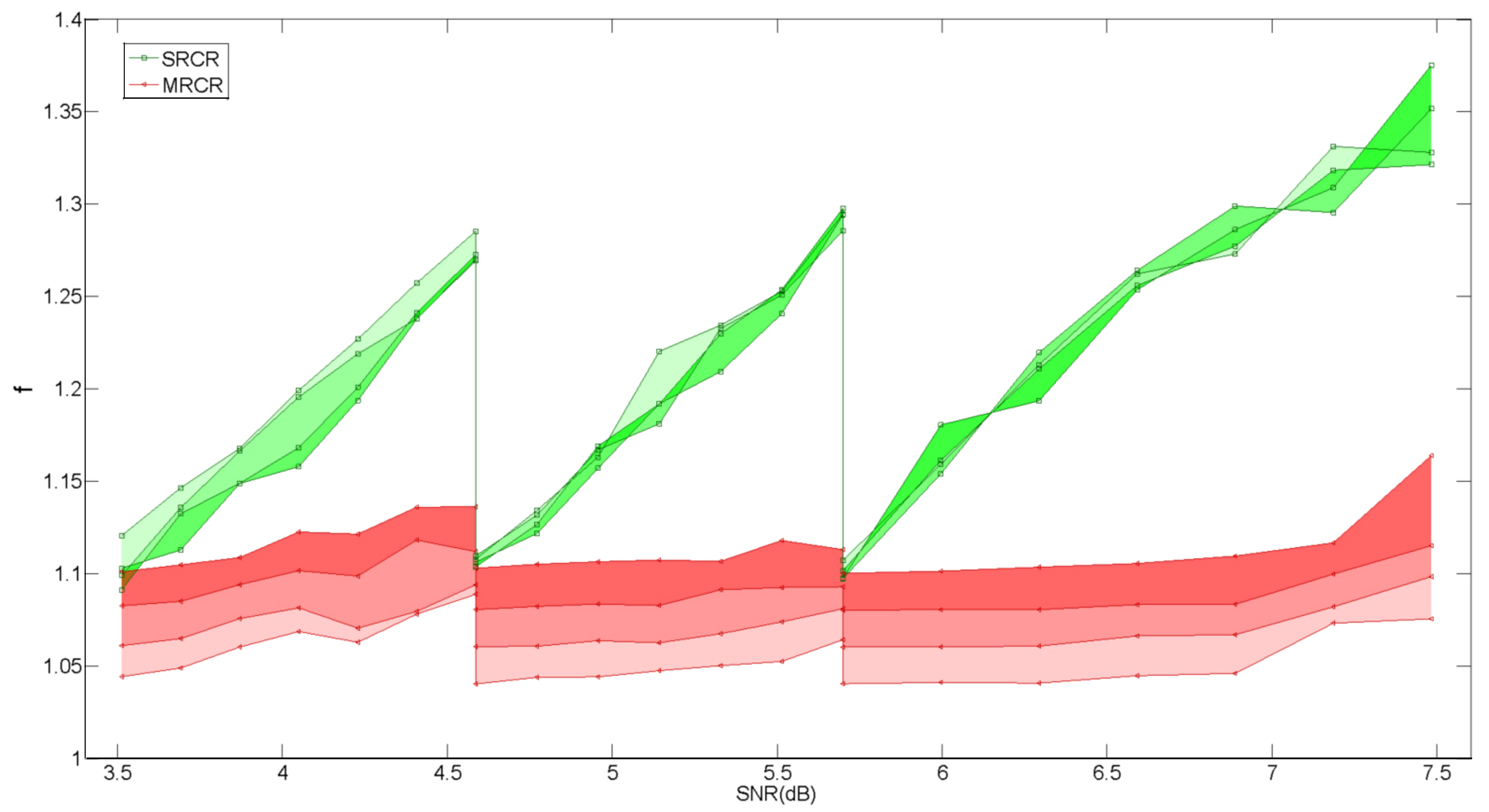}
    \caption{\label{fig:fd} Reconciliation efficiency $f$ of SRCR and MRCR is plotted as a function of SNR
    for given desired reconciliation efficiencies $f_d=1.1$, $1.08$, $1.06$ and $1.04$, respectively.}
\end{figure*}

To measure the potential of SRCR and MRCR for achieving better $f$ as number of punctured bits increases,
an experiment is conducted with $n=10000$ and $f_d=1.1$, $1.08$, $1.06$ and $1.04$.
For every SNR, $500$ sets of keys were decoded four times at different $f_d$ in MRCR and SRCR.

The simulation results are presented in Fig. \ref{fig:fd}.
It is observed that as $f_d$ shifts from $1.1$ to $1.04$ or as number of punctured bits increases,
the corresponding green lines are limited and intertwined within the slim green areas.
In contrast, there are no interactions among the red lines so that they are distributed on the larger areas in red,
and superior values of $f$ are obtained with $f_d$ from $1.1$ to $1.04$.
Such evidence suggests that compared with SRCR, MRCR has greater potential to achieve better $f$.

\subsection{Throughput}
\begin{figure*}[htbp]
    \centering
    \includegraphics[width=15cm]{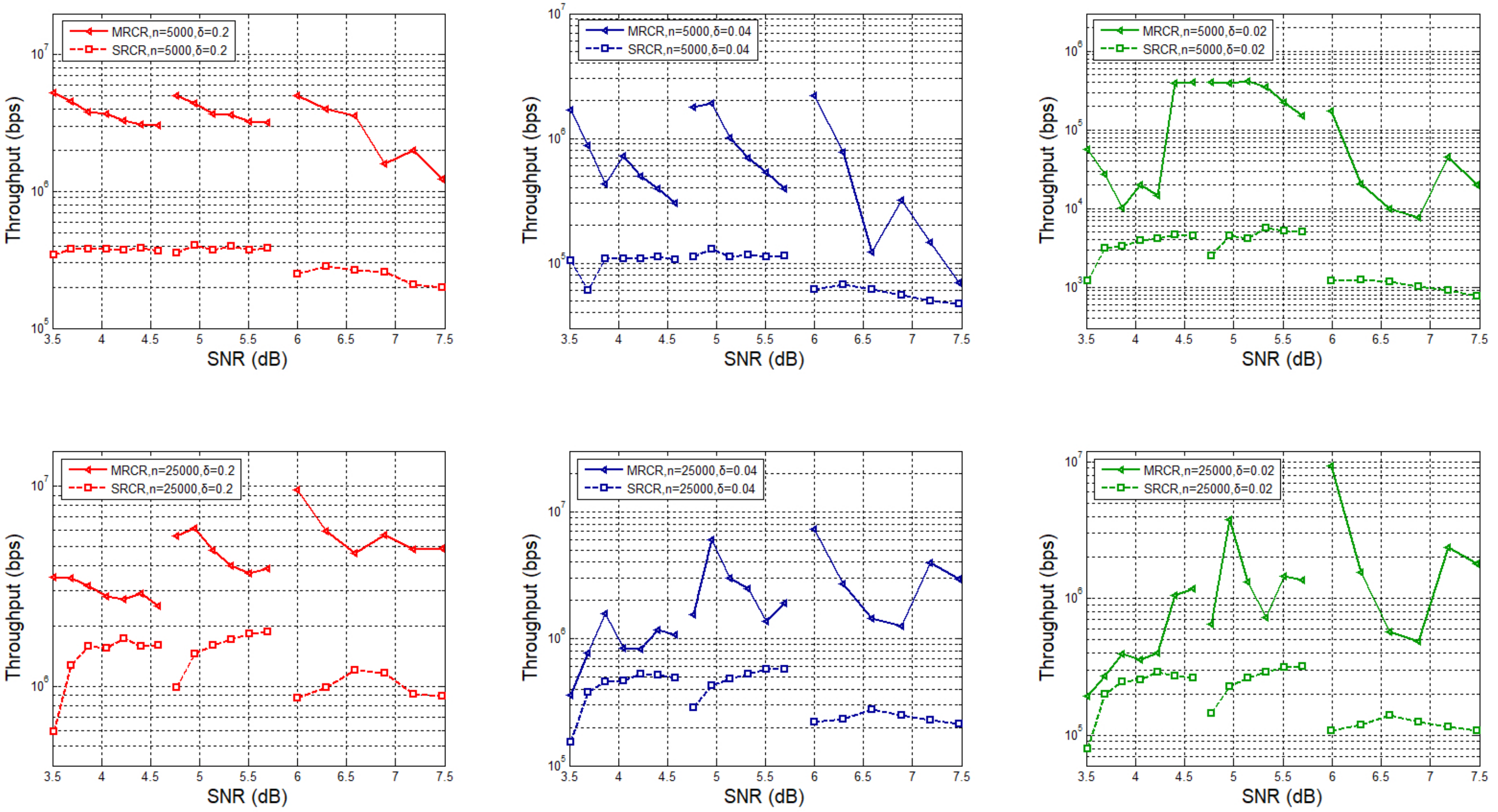}
    \caption{\label{fig:throughput} Throughput of SRCR and MRCR under different SNR values.
    Throughput of MRCR is much higher than that of SRCR for code lengths $n=5000$ and $25000$,
    and $\delta=0.2$, $0.04$ and $0.02$.}
\end{figure*}

\begin{figure*}[htbp]
    \centering
    \includegraphics[width=15cm]{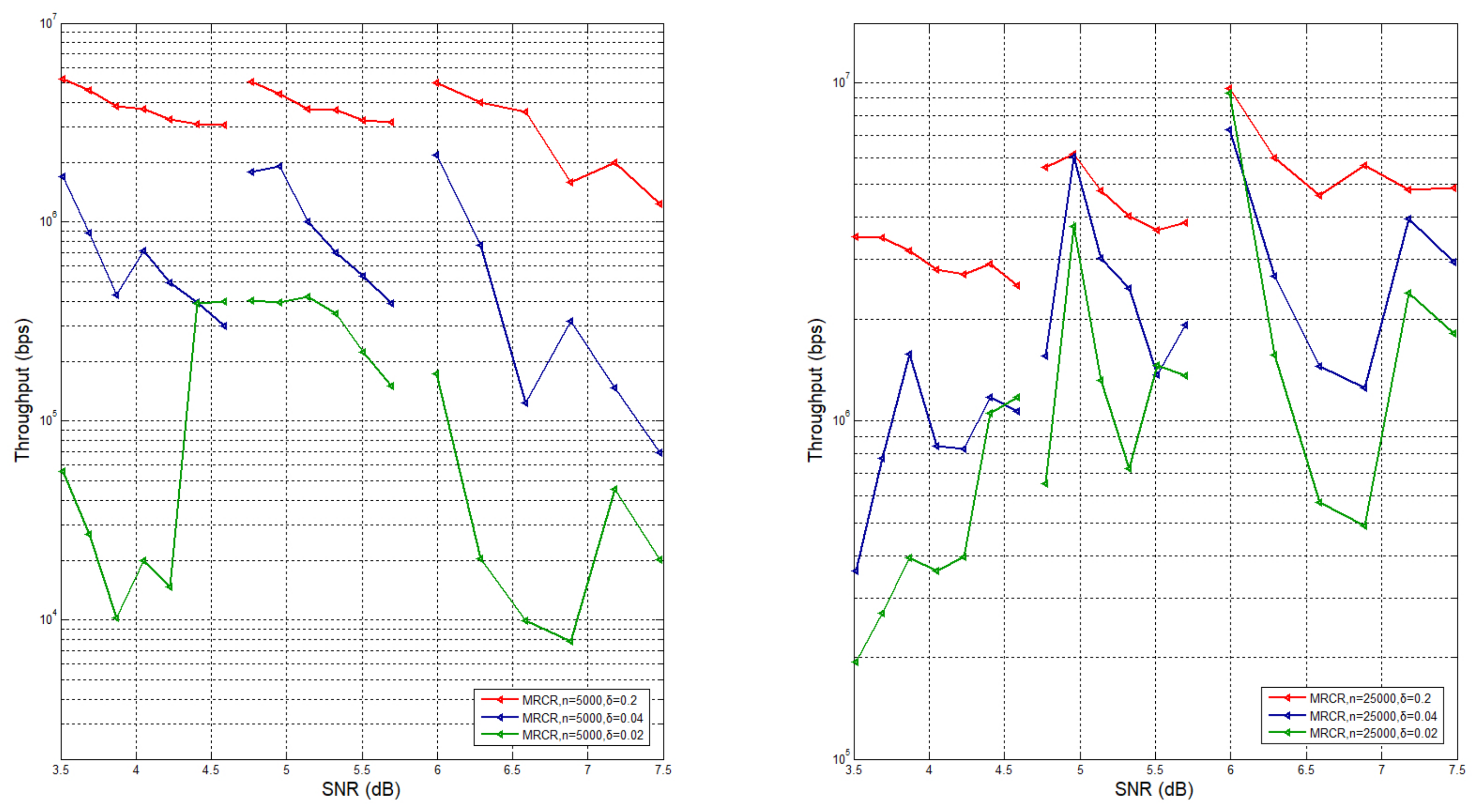}
    \caption{\label{fig:throughput_} Throughput of MRCR under different SNR values.
    The red lines, blue lines and green lines denote the throughput with $\delta=0.2$, $0.04$ and $0.02$, respectively.}
\end{figure*}

As discussed before, MRCR has faster convergence speed, and it can shorten the time by reconciliation in parallel.
So the throughput of MRCR is higher than that of SRCR.
To verify the view, we measure the throughput of SRCR and MRCR in two code lengths, i.e., $n=5000$ and $50000$,
and three $\delta$ values, i.e., $\delta=0.2$, $0.04$ and $0.02$.
Similarly, the range of SNR is divided into three parts, each of which corresponds to one initial code rate $R_0$.
At each SNR, $500$ sets of keys are tested with $fd=1.1$.
And the throughput is calculated as follows,
\begin{equation}
T=N_{success}\cdot(n-P_0)/t,
\label{equ:throughput}
\end{equation}
where $N_{success}$ is the number of successful reconciliation,
and $t$ is the duration of performing these $500$ sets of keys.
The results of throughput are recorded and showed in Fig. \ref{fig:throughput}.
It's clear that the solid lines are much higher than the dotted lines regardless of code lengths, code rates, SNR and $\delta$.
It indicates that MRCR is beneficial to improve the throughput compared with SRCR.
However, because there is only one communication round for SR and MR in one reconciliation,
substantially the throughput of MRCR is less than those of SR and MR.

In addition, we extract solid lines from Fig. \ref{fig:throughput} to form Fig. \ref{fig:throughput_} for further comparisons.
As we can see in Fig. \ref{fig:throughput_}, throughput increases with the increase of $\delta$ values,
the reason of which is further elaborated below.
For a reconciliation process, there is a exact number of shortened bits $\bar{s}$,
at which the reconciliation chances to achieve success and thus obtains optimal $f$.
With the increase of $\delta$,
the number of communication rounds needed to achieve or surpass $\bar{s}$ shortened bits declines,
i.e., the convergence speed increases so that the throughput increases.
However, for lower $\delta$ values,
it is easier to approach $\bar{s}$ to obtain optimal $f$ because of smaller $P2S$ values in Eq \ref{equ:P2S}.
Therefore, there is a balance between throughput and reconciliation efficiency needed to be considered
when deciding $\delta$.


\section{\label{sec:level5}Conclusion}

MR algorithm was proposed to improve the throughput by increasing convergence speed.
In order to improve the reconciliation efficiency with the premise of maintaining the advantage of throughput,
in this paper we introduce the rate-compatible technologies, i.e. shortening and puncturing, into MR.
The numerical results show that MRCR outperforms SRCR at any code length, code rate and SNR.
Moreover, the throughput of MRCR is higher than that of SRCR.
In this regard, MRCR is beneficial to improve secure key generation rate of QKD systems.

\begin{acknowledgments}
This research is financially supported by the National Key Research and Development Program of China (No. 2017YFA0303700),
the Major Program of National Natural Science Foundation of China (No. 11690030, 11690032),
the National Natural Science Foundation of China (No. 61771236),
the Natural Science Foundation of Jiangsu Province (BK20190297)
\end{acknowledgments}

\nocite{*}

\bibliography{apssamp}

\end{document}